\documentclass{pasj02}

\usepackage{amsmath}
\usepackage{xspace}
\usepackage[switch,mathlines]{lineno}
\usepackage{natbib}
\usepackage{xcolor}

\newcommand\cag{${}^{12}{\rm C}(\alpha,\gamma){}^{16}{\rm O}$\xspace}
\newcommand\nickel{$^{56}\mathrm{Ni}$\xspace}

\makeatletter
\DeclareRobustCommand\pasjcite[2]{\ifNAT@swa#2\else#1\fi}
\makeatother

\jyear{2026}
\Received{}
\Accepted{}

\begin{document}

\title{Temperature-resolved sensitivities of $\mathbf{^{56}{Ni}}$ production to helium-burning reactions in pair-instability supernovae}

\author{Hiroki \textsc{Kawashimo}\altaffilmark{1,2}\altemailmark}%
\email{h-kawashimo@g.ecc.u-tokyo.ac.jp}

\author{Nobuya \textsc{Nishimura}\altaffilmark{3,4,5}}%

\author{Yudai \textsc{Suwa}\altaffilmark{1,6}}%

\altaffiltext{1}{Department of Earth Science and Astronomy, Graduate School of Arts and Sciences, The University of Tokyo, 3-8-1 Komaba, Meguro, Tokyo 153-8902, Japan}
\altaffiltext{2}{RIKEN Nishina Center for Accelerator-Based Science, RIKEN, 2-1 Hirosawa, Wako, Saitama 351-0198, Japan}
\altaffiltext{3}{Academic Support Center, Kogakuin University, 2665-1 Nakano-cho, Hachioji, Tokyo 192-0015, Japan}
\altaffiltext{4}{Center for Nuclear Study, The University of Tokyo, 7-3-1 Hongo, Bunkyo, Tokyo 113-0033, Japan}
\altaffiltext{5}{Astrophysical Big-Bang Laboratory, Pioneering Research Institute, RIKEN, 2-1 Hirosawa, Wako, Saitama 351-0198, Japan}
\altaffiltext{6}{Center for Gravitational Physics and Quantum Information, Yukawa Institute for Theoretical Physics (YITP), Kyoto University, Oiwakecho, Sakyo, Kyoto 606-8502, Japan}


\maketitle

\begin{abstract}
We propose a temperature-resolved Monte Carlo (MC) approach to identify the temperature ranges in which low-energy helium-burning reaction rates most strongly affect nucleosynthesis in very massive stars that undergo pair-instability supernovae (PISNe). By performing MC simulations of PISNe, we quantify how temperature-dependent variations in key helium-burning reaction rates, i.e., the triple-$\alpha$ and \cag rates, influence \nickel synthesis. Thousands of stellar evolution calculations using {\tt MESA} reveal that both the \cag and triple-$\alpha$ reactions exhibit their strongest sensitivity at $T \simeq 2.5 \times 10^{8}\,\mathrm{K}$, but with opposite correlation signs. We show that this temperature corresponds to the regime in which the ratio of the sampled rate multipliers is most clearly imprinted on the pre-carbon-burning C/O composition. This demonstrates that PISN nucleosynthesis can probe helium-burning reaction rates in specific low-temperature regimes.
\end{abstract}


\section{Introduction}

Very massive stars (VMSs), with progenitor masses of $\sim200\, M_\odot$, are theoretically predicted to end their evolution as pair-instability supernovae (PISNe) \cite[e.g.,][]{1967PhRvL..18..379B, 2001ApJ...550..372F, 2003ApJ...591..288H}. A PISN is triggered by the creation of electron--positron pairs in the massive C/O core, which reduces the radiation-pressure support. This induces rapid core contraction and explosive oxygen burning, ultimately leading to the complete disruption of the star. Such a thermonuclear explosion produces a large amount of radioactive iron-group nuclei, predominantly \nickel, which serves as the main energy source of the supernova (SN) light curve. As PISNe are expected to be significantly brighter than ordinary SNe, their observational identification is of particular importance. Recently, a possible PISN candidate has been observed \citep{2024A&A...683A.223S}, highlighting the need for further theoretical investigation of the observational properties of PISNe.

Although theoretical models are essential for interpreting PISNe, simulations of PISN progenitor evolution still suffer from large uncertainties in several physical aspects. In addition to the basic modeling of stellar evolution, nuclear reaction rates, in particular, remain significantly uncertain, even for the main nuclear-burning reactions, such as \cag and triple-$\alpha$ \cite[see, e.g.,][]{2009ApJ...702.1068T, 2010ApJ...718..357T, 2009A&A...507.1617D, 2011ApJ...741...61S, 2015PTEP.2015f3E01K, 2023MNRAS.519L..32T, 2023A&A...679A..75T, 2025arXiv250211012X}. Despite extensive efforts over several decades in nuclear physics, determining these rates remains highly challenging both experimentally and theoretically \cite[e.g.,][]{2017RvMP...89c5007D, 2005Natur.433..136F, 2020PhRvL.125r2701K}.

The amount of \nickel synthesized in a PISN is one of the key observables, because it largely determines the brightness of the SN light curve. However, theoretical predictions of the \nickel yield are affected by various physical uncertainties in stellar evolution and explosion models, which complicates a definitive comparison between theory and observations. In particular, previous studies \citep{2018ApJ...863..153T, 2024MNRAS.531.2786K} have shown that the amount of \nickel synthesized in PISNe strongly depends on helium-burning reaction rates. This dependence implies that uncertainties in helium-burning reaction rates directly propagate into uncertainties in the predicted \nickel yield, and hence into the expected luminosity and observability of PISNe.

For a realistic investigation of such reaction-rate uncertainties, the rates should not be treated as uniformly uncertain over the entire temperature range. Throughout stellar evolution and explosion, different temperature ranges may play different roles, and the sensitivity of the final nucleosynthesis products to an individual reaction rate can vary accordingly. Such temperature dependence is also well known in nuclear physics, where reaction cross sections depend strongly on energy and where the dominant physical processes governing the rates may differ between energy ranges. Thus, identifying the temperature ranges in which reaction-rate variations most strongly affect \nickel production is essential both for interpreting future PISN observations and for guiding nuclear cross-section measurements. In previous studies, the \cag and triple-$\alpha$ reaction rates were artificially scaled uniformly up and down over the full temperature range, allowing the authors to estimate upper and lower bounds on \nickel production under such extreme assumptions. However, actual reaction rates are unlikely to deviate so strongly over the entire temperature range. Moreover, such uniform scaling does not identify which temperature ranges are responsible for the resulting change in the final \nickel yield.

During helium burning, the triple-$\alpha$ reaction produces $^{12}\mathrm{C}$, while the \cag reaction converts $^{12}\mathrm{C}$ into $^{16}\mathrm{O}$. The two reactions therefore act as the production and destruction channels of carbon during the formation of the CO core. This makes it necessary to consider both reactions simultaneously when investigating how helium-burning reaction-rate uncertainties affect PISN nucleosynthesis. In this work, we develop a temperature-resolved Monte Carlo (MC) approach to identify the temperature ranges in which helium-burning reaction-rate variations most strongly affect \nickel production in PISNe. Using thousands of stellar evolution models with independently sampled temperature-dependent multiplier factors for the triple-$\alpha$ and \cag reactions, we quantify the statistical relationship between the rate variation at each temperature and the final \nickel yield. We introduce the methodology in section~\ref{sec:method}, present the results in section~\ref{sec:results},
discuss their implications in section~\ref{sec:discussion}, and summarize our conclusions in section~\ref{sec:conclusion}.

\section{Method}\label{sec:method}

\subsection{Stellar evolution calculations with MESA}

We perform stellar evolution simulations of VMSs using the stellar evolution code MESA, version r15140 \citep{2011ApJS..192....3P, 2013ApJS..208....4P, 2015ApJS..220...15P, 2018ApJS..234...34P, 2019ApJS..243...10P, 2023ApJS..265...15J}. The equation of state (EOS) is constructed from several sources: OPAL \citep{Rogers2002}, SCVH \citep{Saumon1995}, FreeEOS \citep{Irwin2012}, HELM \citep{Timmes2000}, and PC \citep{Potekhin2010}. Radiative opacities are primarily taken from OPAL \citep{Iglesias1993, Iglesias1996}, supplemented by low-temperature data from \citet{Ferguson2005} and high-temperature, Compton-scattering-dominated opacities from \citet{Buchler1976}. Electron-conduction opacities are taken from \citet{Cassisi2007}.

Nuclear reaction rates are based on JINA REACLIB \citep{Cyburt2010}, except for the triple-$\alpha$ and \cag reactions, and are supplemented by several weak reaction rate tables \citep{Fuller1985, Oda1994, Langanke2000}. Screening effects are treated according to the prescription of \citet{Chugunov2007}, and thermal neutrino loss rates are adopted from \citet{Itoh1996}.

We adopt a hydrogen-envelope-stripped VMS model with an initial mass of $100\,M_\odot$, following the scenario proposed by \citet{2019ApJ...882...36M}, assuming a metallicity of $Z = 10^{-5}$ for both the stellar components and their environment. We employ a nuclear reaction network based on {\tt approx21\_plus\_co56.net}, which follows the major $\alpha$-chain isotopes and includes radioactive $^{56}\mathrm{Ni}$ and $^{56}\mathrm{Co}$. We note that the additional calculations for SN~2018ibb in section~\ref{sec:sn2018ibb} use different helium-core masses and metallicities, as described there.

Each model is initialized following the standard MESA procedure, where the initial stellar model is constructed as a chemically homogeneous gaseous sphere \citep{2011ApJS..192....3P}. A model is regarded as a successful PISN when all stellar material becomes unbound from the gravitational potential, which means that every mass shell has a velocity exceeding the escape velocity.

\subsection{Temperature-resolved MC sampling of reaction rates}
\label{subsec:mc}
\begin{figure}
    \centering
    \includegraphics[width=80mm]{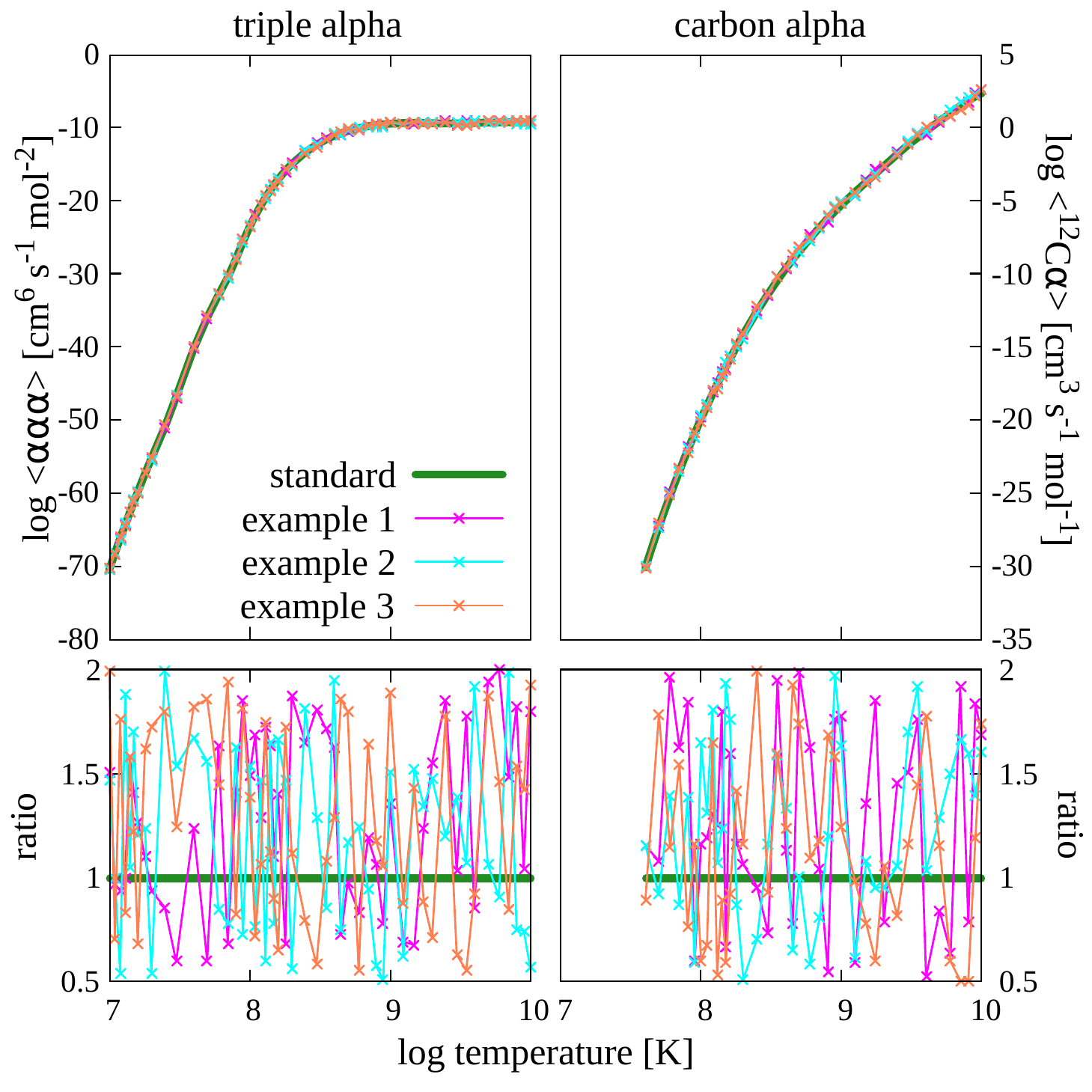}
    \caption{
   Upper: Absolute reaction rates shown as $p^{\mathcal{R}}_i$ in equation~(\ref{eq:ratedif}). $\langle\alpha\alpha\alpha\rangle$ denotes the reaction rate of $3\alpha$, and $\langle^{12}\mathrm{C}\alpha\rangle$ denotes that of \cag. Lower: Reaction rates normalized to the standard {\tt STARLIB} values \citep{2013ApJS..207...18S, 1999NuPhA.656....3A, 2002ApJ...567..643K} shown as $f^{\mathcal{R}}_i$ in equation~(\ref{eq:ratedif}). The green lines show the standard {\tt STARLIB} values, while the other lines show examples of randomly generated reaction rate realizations.
    
    {Alt text: Four panels show the triple-alpha reaction rate on the left and the carbon-alpha reaction rate on the right as functions of logarithmic temperature. In the upper panels, the standard reaction rates are shown by green lines, while three randomly generated rate realizations are shown by magenta, cyan, and orange lines with cross markers for examples 1, 2, and 3, respectively. The lower panels show the corresponding ratios to the standard rates, with the same color coding, varying between approximately 0.5 and 2.}
}
    \label{fig:rand}
\end{figure}

Our goal is to identify the characteristic temperature ranges of the triple-$\alpha$ and \cag reactions that show the strongest sensitivity of the final \nickel yield in the evolution of VMSs leading to PISNe. To this end, we perform thousands of stellar evolution simulations using MC-modified reaction rate tables. By correlating the final \nickel yield with the random multiplier assigned at each temperature, this method identifies the temperature windows in which uncertainties in the helium-burning reaction rates are most strongly propagated to PISN nucleosynthesis. These temperature windows indicate where improved nuclear physics inputs, or future observational constraints on PISN nucleosynthesis, may help identify the rate inputs most relevant to PISN predictions. This differs from studies that perform MC analyses by simultaneously varying multiple reaction rates involved in nucleosynthesis \citep[e.g.,][]{2026MNRAS.546ag152N}, where the rate-uncertainty factors are assumed to vary uniformly over the entire temperature range.

To evaluate the impact of reaction-rate uncertainties on nucleosynthesis, we generated MC-sampled tables based on {\tt STARLIB} \citep{2013ApJS..207...18S}. We adopt the rate by \citet{1999NuPhA.656....3A} for the triple-$\alpha$ reaction and the rate by \citet{2002ApJ...567..643K} for the \cag reaction. For each stellar model $i$, each reaction channel $\mathcal{R}$ (e.g., triple-$\alpha$ or \cag reaction), and each temperature point $T$, a multiplier $f_i^{\mathcal{R}}(T)$ is sampled from a uniform distribution over $[0.5,2.0]$, thereby introducing distinct temperature-dependent variations in the reaction rates. In this work, we apply this procedure to the triple-$\alpha$ and \cag reactions, which are both critical during the same evolutionary stage. The random multipliers for these two reactions are generated simultaneously in each model. This approach allows us to investigate whether, and to what extent, competition between the triple-$\alpha$ and \cag reactions affects \nickel synthesis.
 
The multiplicative factors were drawn from a uniform distribution over $[0.5,2.0]$, corresponding to a variation within a factor of two of the recommended rate. This range was adopted as a simple exploratory envelope, comparable to the scale of the rate uncertainties relevant to the helium-burning reactions considered here.\footnote{For example, in the \cag rate of \citet{2002ApJ...567..643K},
the ratio of the high rate to the adopted rate reaches at most $\simeq 1.37$,
while the ratio of the low rate to the adopted rate reaches at least
$\simeq 0.69$ over $0.04<T_9<10$.} We emphasize that the resulting rate tables are not intended to represent statistically sampled realizations of the {\tt STARLIB} rate uncertainties. In {\tt STARLIB}, rate uncertainties are given as temperature-dependent log-normal factors, and realistic nuclear uncertainties are generally correlated across temperature. Our goal is instead to perform a temperature-resolved sensitivity analysis: by perturbing the rate at each temperature point independently, we identify the temperature ranges in which variations of the helium-burning rates are most strongly associated with changes in the final \nickel yield.

Figure~\ref{fig:rand} shows illustrative examples of the randomly generated \cag\ and triple-$\alpha$ reaction rate tables. The green line indicates the standard rate, while the colored lines represent different MC realizations. This demonstrates the extent of the random variations introduced in our analysis.

The perturbed reaction rates for model $i$ are then calculated as
\begin{equation}
p_i^{\mathcal{R}}(T)
=
p_{\mathrm{std}}^{\mathcal{R}}(T)\times
f_i^{\mathcal{R}}(T).
\label{eq:ratedif}
\end{equation}
Subsequent nucleosynthesis calculations are performed for each model to obtain the synthesized \nickel mass at the end of the calculation, denoted by $M_{^{56}\mathrm{Ni},i}$. For each reaction $\mathcal{R}$ and temperature $T$, we investigate the relationship between $M_{^{56}\mathrm{Ni},i}$ and the corresponding sampled multiplier $f_i^{\mathcal{R}}(T)$ and quantify their statistical correlation using the Pearson correlation coefficient:
\begin{equation}
\label{eq:corr}
r^{\mathcal{R}}(T)
=
\frac{
\sum\limits_{i=1}^{n}
\left( M_{^{56}\mathrm{Ni},i} - \overline{M_{^{56}\mathrm{Ni}}} \right)
\left( f_i^{\mathcal{R}}(T) - \overline{f^{\mathcal{R}}(T)} \right)
}{
\sqrt{
\sum\limits_{i=1}^{n}
\left( M_{^{56}\mathrm{Ni},i} - \overline{M_{^{56}\mathrm{Ni}}} \right)^2
}
\sqrt{
\sum\limits_{i=1}^{n}
\left( f_i^{\mathcal{R}}(T) - \overline{f^{\mathcal{R}}(T)} \right)^2
}
}.
\end{equation}
Here, $n$ denotes the number of stellar models included in the correlation calculation, namely models that successfully produced a final \nickel yield. Models that failed to explode or whose calculations did not complete successfully account for less than about 10\% of the total attempted sample and are excluded (944 failed models in 10,000 trials in our work) from both $n$ and the sums in equation~(\ref{eq:corr}). The overbar indicates the average over the models included in the correlation calculation. This procedure provides a sensitivity analysis that identifies the temperature windows in which each reaction shows the strongest linear association with \nickel synthesis. To satisfy detailed balance, the same factor is also applied to the reverse reaction rates.

\section{Results}
\label{sec:results}
\subsection{Temperature-dependent sensitivity of key reactions}\label{sec:main_result}

\begin{figure}[t]
    \centering
    \includegraphics[width=80mm]{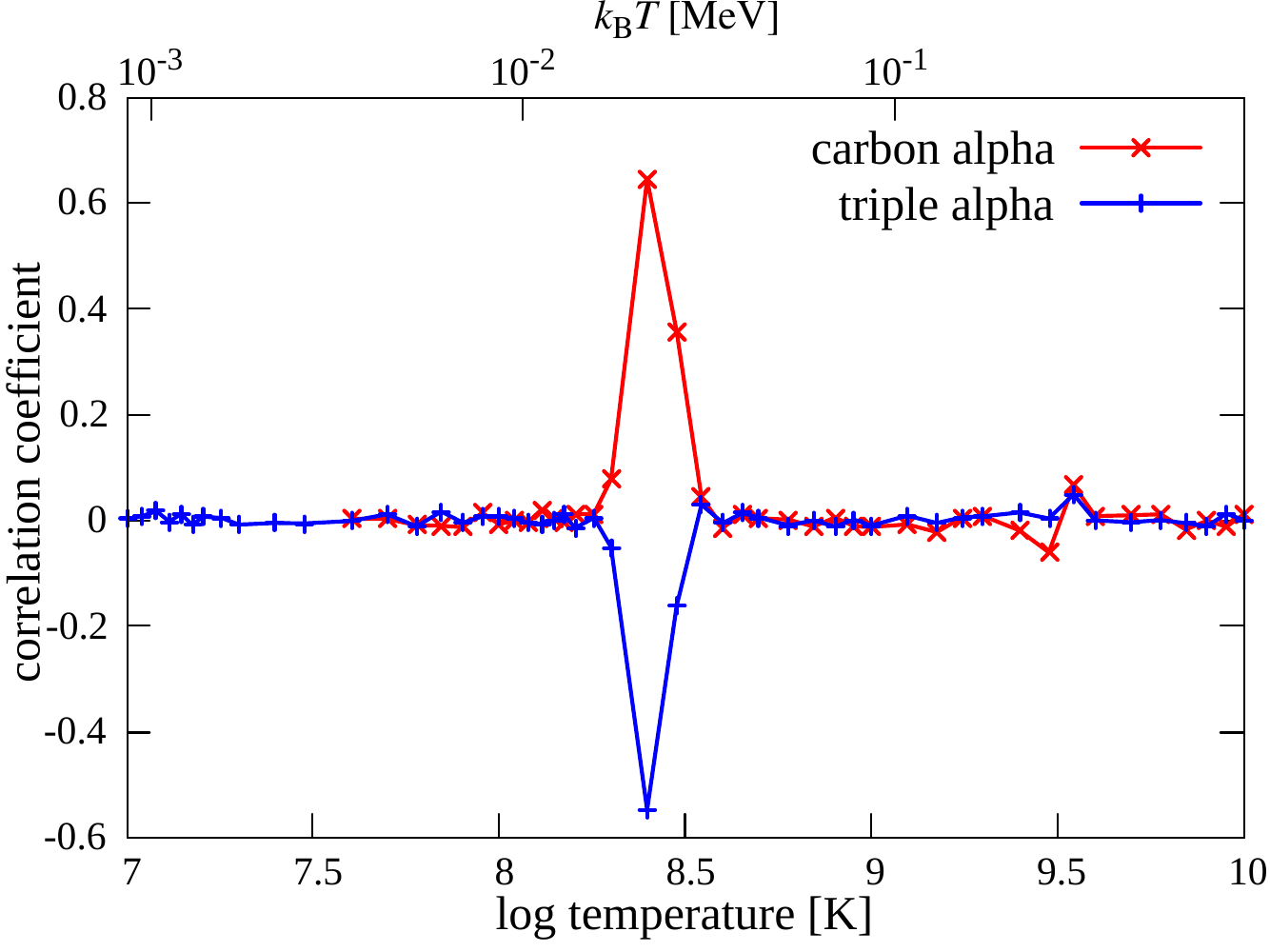}
    \caption{
    Temperature dependence of the Pearson correlation coefficient between the temperature-dependent rate multipliers and the synthesized \nickel mass. The solid curves represent $r^{3\alpha}(T)$ (blue) and $r^{\mathrm{C}\alpha}(T)$ (red). Positive values indicate a direct correlation between rate enhancement and \nickel production, while negative values indicate an inverse correlation.
    {Alt text: Correlation coefficient on the vertical axis as a function of logarithmic temperature in kelvin on the horizontal axis. The carbon-alpha curve (red line) has a sharp positive peak near $2.5\times10^8$ K, while the triple-alpha curve (blue line) has a negative minimum near the same temperature; both remain close to zero over most other temperatures.}
}
    
    \label{fig:result}
\end{figure}

Using the temperature-resolved MC sampling described in section~\ref{sec:method}, we performed 10,000 stellar evolution calculations with independently sampled rate multipliers.

Figure~\ref{fig:result} presents the temperature dependence of the Pearson correlation coefficient between the sampled multiplier and the final \nickel yield for the triple-$\alpha$ and \cag reactions. The vertical axis shows the correlation coefficient, while the horizontal axis corresponds to temperature. The blue and red curves show $r^{3\alpha}(T)$ for the triple-$\alpha$ reaction and $r^{\mathrm{C}\alpha}(T)$ for the \cag reactions, respectively.

Both reactions exhibit pronounced sensitivity around $2.5 \times 10^8\,\mathrm{K}$, yet with opposite signs: \cag is associated with a strong positive correlation ($r^{\mathrm{C}\alpha}(2.5 \times 10^8\,\mathrm{K})\simeq0.644$), whereas the triple-$\alpha$ reaction shows a negative correlation ($r^{3\alpha}(2.5 \times 10^8\,\mathrm{K})\simeq-0.548$). This contrast suggests that the two reactions affect \nickel synthesis in opposite directions within the same temperature range.
Replacing $M_{^{56}\mathrm{Ni},i}$ in equation~(\ref{eq:corr}) with the final explosion energy, we find that the strongest correlations again occur at $T\simeq2.5\times10^{8}\,\mathrm{K}$, with Pearson coefficients of $-0.604$ for the triple-$\alpha$ reaction and $0.641$ for the \cag reaction.

The sign difference arises from their distinct nuclear roles. The triple-$\alpha$ reaction facilitates the production of ${}^{12}\mathrm{C}$ by consuming $\alpha$-particles during helium burning, while \cag converts ${}^{12}\mathrm{C}$ into ${}^{16}\mathrm{O}$, thereby reducing the available carbon. As a result, the two reactions act in competition, and their balance influences the pre-explosive composition of the CO core.

The $3\alpha$ reaction contributes to the growth of the CO core. Since heavier CO cores are generally associated with stronger explosions and increased \nickel synthesis \citep{2016MNRAS.456.1320T}, this reaction might be expected to enhance \nickel production. However, the negative correlation observed here suggests that, within this temperature range, the suppressive effect of increased carbon abundance on oxygen burning dominates over the indirect effect of core mass growth. This balance highlights the utility of correlation analysis in disentangling the relative importance of such competing nuclear effects.

These results indicate that the \nickel yield is sensitive not only to the explosive oxygen-burning phase itself, but also to the progenitor composition established during helium burning. Variations in the helium-burning reaction rates around $T\simeq2.5\times10^8\,\mathrm{K}$ can therefore affect the final \nickel yield indirectly, by modifying the pre-explosive C/O composition. In the next subsection, we examine this compositional imprint directly.

\subsection{Imprint of helium-burning rates on the pre-explosive C/O composition}
\label{subsec:co}
\begin{figure}[t]
    \centering
    \includegraphics[width=80mm]{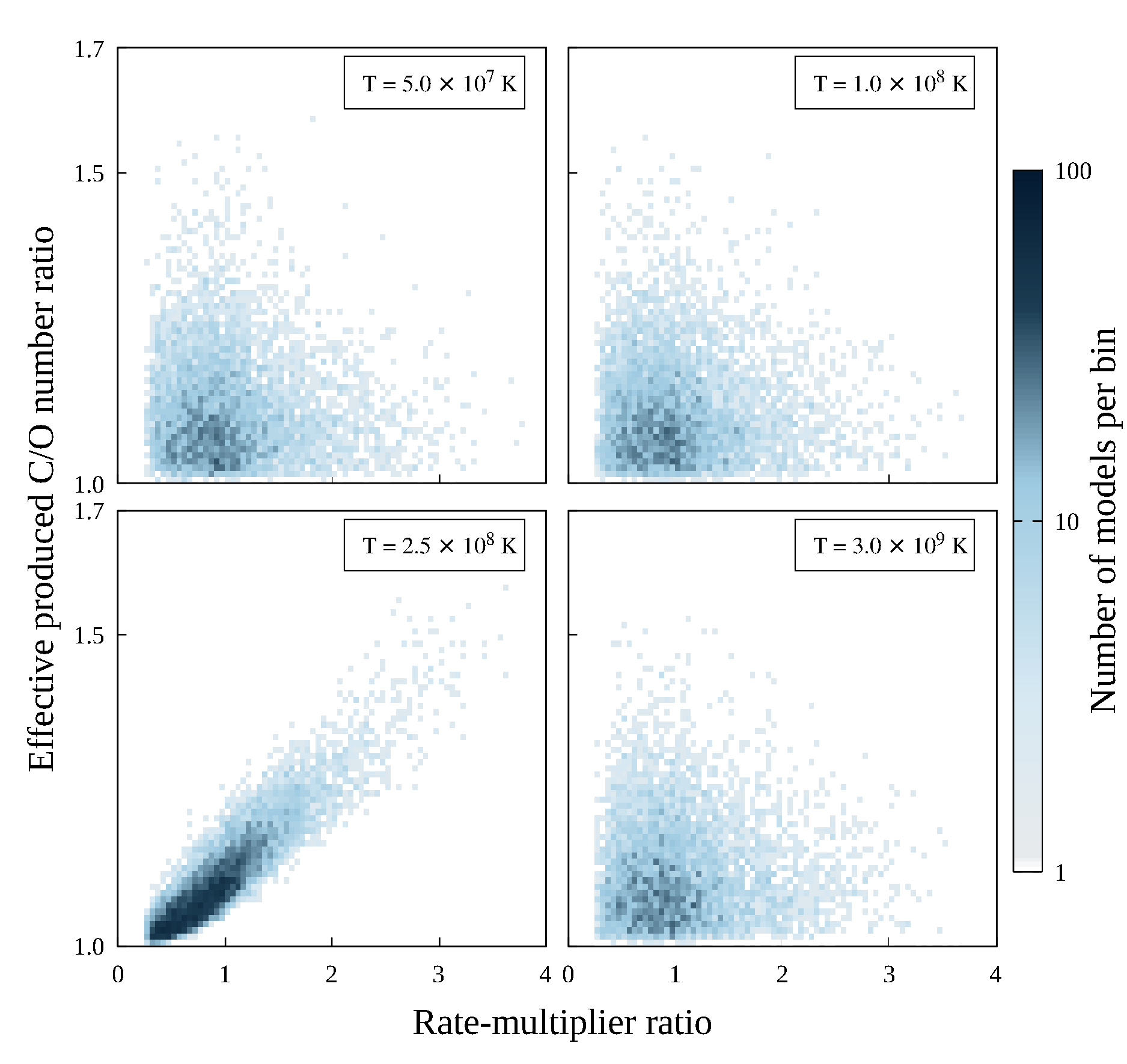}
    \caption{Two-dimensional histograms showing the relation between the rate-multiplier ratio and the effective C/O number ratio of produced nuclei at $5.0\times10^{7}$, $1.0\times10^{8}$, $2.5\times10^{8}$, and $3.0\times10^{9}\,\mathrm{K}$. The horizontal axis indicates the rate-multiplier ratio, $\mathcal{F}_i(T)$, defined in equation~(\ref{eq:mathcalF}). The vertical axis indicates the effective produced C/O number ratio, evaluated from the total stellar species masses at $T_{\mathrm{c}}=10^{8.7}\,\mathrm{K}$, before the onset of carbon burning. All panels are plotted with the same binning, axis ranges, and color scale. The color indicates the number of models per bin, shown on a logarithmic scale. A distinct structured distribution is seen only at $2.5\times10^{8}\,\mathrm{K}$, whereas the distributions at the other temperatures are comparatively diffuse.

    {Alt text: Four two-dimensional histograms show the rate-multiplier ratio on the horizontal axis and the effective produced C/O number ratio on the vertical axis at $5.0\times10^7$, $1.0\times10^8$, $2.5\times10^8$, and $3.0\times10^9$ K. The color scale gives the number of models per bin on a logarithmic scale. The $2.5\times10^8$ K panel shows a narrow positively sloped distribution, whereas the distributions in the other three panels are broader and less structured.}
}
    
    \label{fig:dist}
\end{figure}

The temperature sensitivity identified in figure~\ref{fig:result} suggests that helium-burning reaction rates leave their strongest imprint on the progenitor composition around $T\simeq2.5\times10^8\,\mathrm{K}$. This imprint can be understood through the carbon--oxygen composition established before carbon burning. The triple-$\alpha$ reaction first produces ${}^{12}\mathrm{C}$ and initiates the formation of the CO core, while the subsequent \cag reaction consumes ${}^{12}\mathrm{C}$ and produces ${}^{16}\mathrm{O}$. Therefore, the balance between these two reaction rates controls how much carbon remains after helium burning and how much material is converted into oxygen and heavier $\alpha$-chain nuclei.

Figure~\ref{fig:dist} shows two-dimensional histograms of the relation between the rate-multiplier ratio and the effective number ratio of produced carbon to produced oxygen. For each model $i$, the rate-multiplier ratio at temperature $T$ is defined as
\begin{equation}
\mathcal{F}_i(T)
=
\frac{
f_i^{3\alpha}(T)
}{
f_i^{\mathrm{C}\alpha}(T)
}.
\label{eq:mathcalF}
\end{equation}
Here, $f_i^{3\alpha}(T)$ and $f_i^{\mathrm{C}\alpha}(T)$ are the sampled multipliers applied to the standard triple-$\alpha$ and \cag rates, respectively. The effective number ratio of produced carbon to produced oxygen is evaluated from the total stellar species masses when the central temperature reaches $T_{\mathrm{c}}=10^{8.7}\,\mathrm{K}$. This temperature is chosen because central helium burning has almost ended, whereas carbon burning has not yet started. Thus, apart from the negligible contribution of the initial metallicity, nuclei heavier than carbon can be regarded as products of successive $\alpha$ captures. We define the effective produced C/O number ratio as
\begin{equation}
\label{eq:effectiveco}
\frac{
M(^{12}\mathrm{C})
+\frac{3}{4}M(^{16}\mathrm{O})
+\frac{3}{5}M(^{20}\mathrm{Ne})
+\frac{3}{6}M(^{24}\mathrm{Mg})
}{
\frac{3}{4}\times\left[
M(^{16}\mathrm{O})
+\frac{4}{5}M(^{20}\mathrm{Ne})
+\frac{4}{6}M(^{24}\mathrm{Mg})
\right]
},
\end{equation}
where $M(X)$ denotes the total stellar mass of isotope $X$ at $T_{\mathrm{c}}=10^{8.7}\,\mathrm{K}$. The terms proportional to $M(^{16}\mathrm{O})$, $M(^{20}\mathrm{Ne})$, and $M(^{24}\mathrm{Mg})$ in the numerator represent the amount of carbon that must have been produced before subsequent $\alpha$ captures. Similarly, the terms proportional to $M(^{20}\mathrm{Ne})$ and $M(^{24}\mathrm{Mg})$ in the denominator represent the corresponding oxygen contribution. Since $(M_{^{12}\mathrm{C}}/12)/(M_{^{16}\mathrm{O}}/16)=(4/3)(M_{^{12}\mathrm{C}}/M_{^{16}\mathrm{O}})$, the final factor of $3/4$ converts the mass ratio into a number ratio. We note that contributions from heavier $\alpha$-chain nuclei are negligible at this stage. The total stellar mass of the next $\alpha$-chain isotope, $^{28}\mathrm{Si}$, remains below $10^{-2}\,M_\odot$ in all models at $T_{\mathrm{c}}=10^{8.7}\,\mathrm{K}$, indicating that the $\alpha$-chain has hardly proceeded beyond $^{24}\mathrm{Mg}$.

Among the temperatures shown in figure~\ref{fig:dist}, only $T\simeq2.5\times10^8\,\mathrm{K}$ exhibits a clearly structured distribution. This indicates that $\mathcal{F}_i(T)$ is most directly imprinted on the C/O composition before carbon burning at this temperature. In contrast, the distributions at the other temperatures are much more diffuse, suggesting that rate variations at those temperatures do not control the pre-carbon-burning C/O composition in the same systematic way. This result provides a physical interpretation of the correlations shown in figure~\ref{fig:result}.

The opposite signs of the correlations in figure~\ref{fig:result} can therefore be understood as a consequence of how the two reactions regulate the residual carbon abundance. Previous studies have shown that the amount of carbon remaining after helium burning is important for determining the strength of the subsequent PISN explosion and the synthesized \nickel mass \citep{2018ApJ...863..153T,2024MNRAS.531.2786K}. A larger \cag rate at $T\simeq2.5\times10^8\,\mathrm{K}$ shifts the helium-burning products toward oxygen-rich compositions, leaving less carbon available for pre-explosive carbon burning. The progenitor then retains a more compact structure until oxygen ignition, resulting in more violent oxygen burning and a larger \nickel yield. In contrast, a larger triple-$\alpha$ rate shifts the composition toward carbon-rich models, where enhanced carbon burning injects energy before oxygen ignition, affects the stellar structure, and suppresses subsequent \nickel production \citep{2024MNRAS.531.2786K}. Our results therefore support the interpretation that the temperature at which the rate-multiplier ratio most clearly determines the pre-carbon-burning C/O composition is also the temperature at which the final \nickel yield becomes most sensitive to the helium-burning reaction rates.

\section{Discussion}
\label{sec:discussion}

\subsection{Convergence}

To validate the reliability of the computed results, we examined how the correlation values evolve with the number of stellar models. This analysis also provides insight into the computational demand associated with our method.

\begin{figure}[t]
    \centering
    \includegraphics[width=80mm]{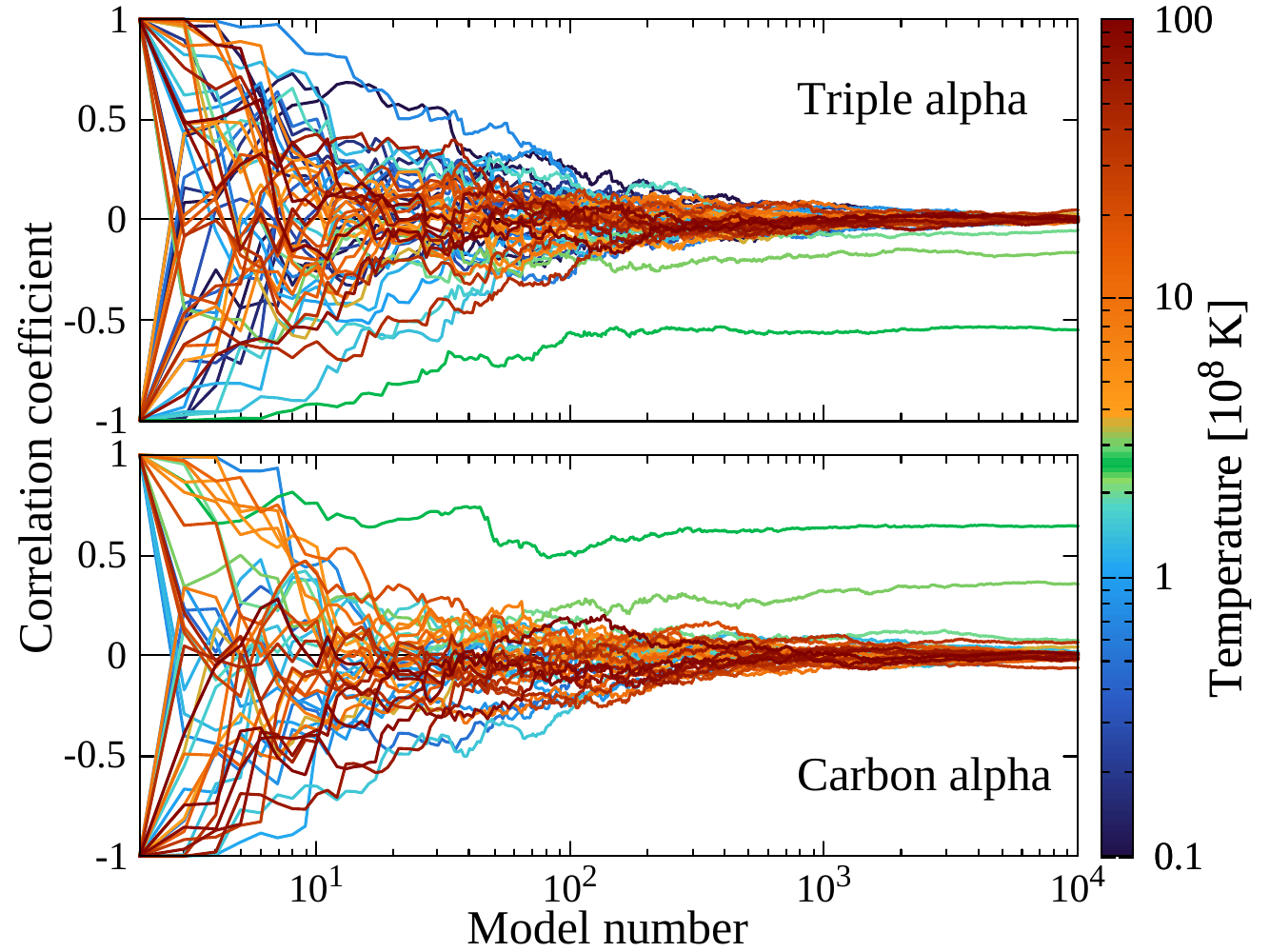}

    \caption{Stepwise evolution of the correlation coefficients as the cumulative number of attempted stellar models is increased. At each step, the correlation coefficient is computed using only the successful models among the attempted models up to that point. Each line corresponds to a different temperature, as indicated by the color scale. The upper panel shows the $3\alpha$ series, and the lower panel shows the \cag series. 
    {Alt text: Two stacked panels show the correlation coefficient on the vertical axis as a function of the cumulative model number on the logarithmic horizontal axis, for the triple-alpha reaction in the upper panel and the carbon-alpha reaction in the lower panel. Curve color represents temperature according to the logarithmic color scale at right, ranging from $0.1\times10^8$ to $100\times10^8$ K. The curves fluctuate widely at small model numbers. At large model numbers, most curves lie close to zero, while the green curves corresponding to temperatures around $2.5\times10^8$ K remain clearly offset from zero, negatively for the triple-alpha reaction and positively for the carbon-alpha reaction.}
}

    \label{fig:conv}
\end{figure}

Figure~\ref{fig:conv} presents the evolution of the correlation coefficient as a function of the cumulative number of attempted stellar models, shown on the horizontal axis. At each point, the correlation coefficient is evaluated using only the successful models among the attempted models up to that point, as in equation~(\ref{eq:corr}). The vertical axis shows the resulting correlation coefficient. Each curve is color-coded by the corresponding temperature, with blue and red representing lower and higher temperatures, respectively. Note that the model count includes cases that did not result in successful explosions. Since such failed-explosion models account for less than about 10\% of the total sample, excluding them would change the effective model count only by a factor of order unity and would not qualitatively affect the convergence behavior.

Candidate temperature windows associated with the dominant correlation signals already emerge after $\mathcal{O}(10^2)$ attempted models. This indicates that the method can efficiently identify promising temperature ranges where reaction-rate variations may strongly affect \nickel synthesis, although larger samples are required for robust estimates of the correlation amplitudes.

\subsection{SN 2018ibb}
\label{sec:sn2018ibb}

SN~2018ibb is one of the most promising PISN candidates, with long-term observational monitoring extending beyond 1000 days \citep{2024A&A...683A.223S}. Several theoretical studies have also supported its interpretation as a PISN (e.g., \citealp{2024A&A...689A..60K}; but see also \citealp{2024ApJ...972...11N}). As an illustrative application of the temperature sensitivity identified above, we discuss the implications for SN~2018ibb under the PISN interpretation.

For this application, we perform additional stellar-evolution calculations separately from the temperature-resolved MC models described in section~\ref{subsec:mc}. For each reaction, the tabulated rate at $T=2.5\times10^8\,\mathrm{K}$ is independently varied from 0.1 to 10 times its standard value, while the tabulated rates at all other grid points are kept at their standard values. The neighbouring grid points at $2.0$ and $3.0\times10^8\,\mathrm{K}$ thus retain their standard values, and MESA evaluates the rate between tabulated points using piecewise monotonic cubic interpolation \citep{2011ApJS..192....3P, 2013ApJS..208....4P}. This wide parameter range is used only to visualize the model response over an extended parameter space and not as a realistic uncertainty range.

\begin{figure*}[h]
    \centering
    \includegraphics[width=160mm]{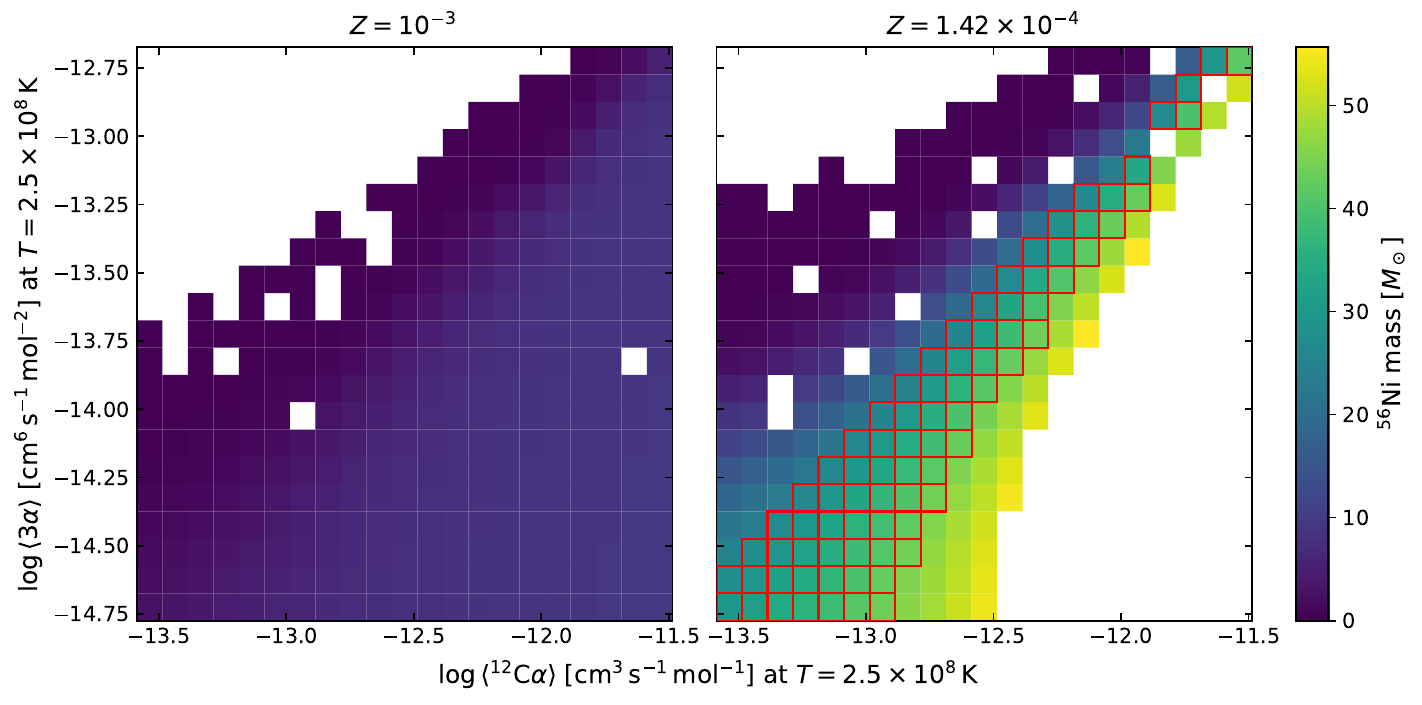}
    \caption{Synthesized \nickel\ mass as a function of the \cag\ reaction rate (horizontal axis) and the triple-$\alpha$ reaction rate (vertical axis) at $T=2.5\times10^{8}\,\mathrm{K}$. The left and right panels show $130\,M_\odot$ helium-core models with $Z=10^{-3}$ and $Z=1.42\times10^{-4}\;(=10^{-2}Z_\odot)$, respectively. The color map shows the synthesized \nickel\ mass for each combination of reaction rates. The red boxes indicate the grid cells in the parameter space that are consistent with the \nickel\ mass inferred for SN~2018ibb, $25$--$44\, M_\odot$. For the white regions in the upper-left and lower-right parts of the panels, see footnote~\ref{fn}.
    
    {Alt text: Two heat maps show synthesized $^{56}\mathrm{Ni}$ mass as a function of the carbon-alpha reaction rate on the horizontal axis and the triple-alpha reaction rate on the vertical axis at $2.5\times10^8$ K, for metallicities $Z=10^{-3}$ on the left and $Z=1.42\times10^{-4}$ on the right. Color indicates the synthesized $^{56}\mathrm{Ni}$ mass. Red outlines, which appear only in the right panel, mark grid cells corresponding to $25$--$44\,M_\odot$ of $^{56}\mathrm{Ni}$, while white regions indicate models without successful PISN explosions.}
}
    
    \label{fig:ibb}
\end{figure*}

Figure~\ref{fig:ibb} shows the synthesized \nickel\ mass for $130\,M_\odot$ helium-core models in the plane of the \cag\ and triple-$\alpha$ reaction rates at $T=2.5\times10^8\,\mathrm{K}$. The red outlined grid cells in the right panel indicate model combinations whose synthesized \nickel\ masses fall within the observationally inferred range for SN~2018ibb, $\sim25$--$44\,M_\odot$. SN~2018ibb is thought to have originated from the explosion of a progenitor with a helium-core mass of $\sim120$--$130\,M_\odot$ \citep{2024A&A...683A.223S}. The left panel assumes the metallicity at which PISNe are typically expected to occur, $Z\sim10^{-3}$ \citep{2024MNRAS.534..151G,2025A&A...703A.215S}. Even when both reaction rates are varied over the wide range of 0.1--10 times their standard values, the synthesized \nickel\ mass does not reach this observationally inferred range in the $Z\sim10^{-3}$ model.\footnote{
White regions indicate models that do not undergo successful PISN explosions. In the carbon-rich upper-left region, convective C-shell burning can lower the core entropy, allowing the star to avoid complete pair-instability disruption \citep{2018ApJ...863..153T,2020ApJ...902L..36F}. In the carbon-poor lower-right region, models reach conditions where photodisintegration absorbs much of the energy released by explosive burning, preventing the star from becoming fully unbound \citep{2024MNRAS.531.2786K}.\label{fn}} Note that the weak color variation in the left panel partly reflects the use of the same linear color scale in both panels, since the synthesized \nickel masses there are generally below $10\,M_\odot$ and variations in this low-yield regime are more clearly seen on a logarithmic scale (see \citealt{2024MNRAS.531.2786K}); the weak visual contrast alone therefore does not imply a shift of the characteristic temperature.

The right panel of figure~\ref{fig:ibb} shows the corresponding result for a lower-metallicity model with $Z=1.42\times10^{-4}\;(=10^{-2}Z_\odot)$. In this case, a region of parameter space appears that is consistent with the \nickel\ mass inferred for SN~2018ibb. Since the outcome of a PISN depends sensitively on the progenitor mass and is therefore strongly affected by metallicity-dependent mass loss, this comparison suggests that SN~2018ibb may have originated from an unusually metal-poor environment, relative to the metallicities expected for the bulk of the PISN population. Although this discussion is only illustrative, it shows that the temperature-resolved reaction-rate sensitivity identified in this work can provide a useful framework for interpreting future PISN observations.

\section{Conclusion}
\label{sec:conclusion}
We conducted large-scale MC stellar evolution simulations to identify the temperature ranges in which helium-burning reaction rates most strongly affect \nickel production in PISNe. The \cag and triple-$\alpha$ reactions show peak sensitivity around $T\simeq2.5\times10^8\,\mathrm{K}$, with opposite correlation signs reflecting their competitive roles in setting the pre-explosive C/O composition. We showed that the rate-multiplier ratio at this temperature is most clearly imprinted on the effective C/O composition before carbon burning, explaining why localized variations in low-temperature helium-burning rates can significantly alter the final \nickel yield. In nuclear-energy terms, the standard Gamow-window expressions (e.g., \citealt{2017RvMP...89c5007D}) give $E_0\simeq370\,\mathrm{keV}$ with $\Delta\simeq210\,\mathrm{keV}$ for \cag at this temperature, while the triple-$\alpha$ reaction proceeds through the ${}^8\mathrm{Be}$ ground-state and Hoyle-state resonances at approximately $92$ and $287\,\mathrm{keV}$ above the respective thresholds \citep{2016PrPNP..89...56B}, placing the identified sensitivity in the few-hundred-keV regime relevant to stellar helium burning.

Our convergence analysis further shows that candidate temperature windows for the dominant sensitivities can already emerge after only $\mathcal{O}(10^2)$ stellar-evolution calculations, although larger samples are required to determine the correlation amplitudes robustly. This rapid emergence of the sensitivity signal is an important advantage of the present temperature-resolved MC approach and makes it practical to identify the key temperature windows of uncertain nuclear reaction rates with a relatively small number of models.

As future wide-field surveys increase the number of PISN candidates, statistical comparisons with theoretical models may provide an observational route to probing low-energy helium-burning reaction rates.
Such constraints would be particularly valuable given the remaining nuclear-physics uncertainties. The triple-$\alpha$ reaction has been extensively studied both experimentally and theoretically, with recent precision measurements of the Hoyle-state radiative decay properties further refining the reaction rate \citep[e.g.,][]{2005Natur.433..136F, 2020PhRvL.125r2701K, 2021PhLB..81736283T, 2025PhLB..87039893S}. Despite this progress, the rate is still less tightly constrained at the lower temperatures relevant to progenitor evolution. The situation is more severe for the \cag\ reaction, whose rate remains difficult to determine directly at the relevant low energies \citep[e.g.,][]{2017RvMP...89c5007D, 2023ApJ...945...41S, 2024PhRvC.109d5808N, 2024PhRvC.110c5809P, 2024PhRvC.110e5803M}, leaving substantial uncertainties. The present temperature-resolved MC framework can also be extended to more complex nucleosynthesis systems involving many coupled reactions, providing a systematic way to connect temperature-dependent nuclear physics uncertainties with observable nucleosynthesis signatures.

The robustness of the exact sensitivity peak against variations in progenitor properties and stellar-evolution physics remains to be systematically explored. Nevertheless, our results suggest that the sensitivity around $T\simeq2.5\times10^8\,\mathrm{K}$ persists within standard evolutionary scenarios: the $130\,M_\odot$ helium-core models in figure~\ref{fig:ibb} remain clearly sensitive to rate variations at this temperature, and previous studies find ordinary variations in metallicity, internal mixing, and stellar winds to have smaller effects on pair-instability outcomes than uncertainties in key nuclear reaction rates \citep{2019ApJ...887...53F,2026arXiv260631757S}. Qualitatively different evolutionary histories may lead to shifts in the exact peak; for example, the triple-$\alpha$ reaction can already operate during hydrogen burning in Population~III massive stars \citep{2009ApJ...706.1184O}. Additional nuclear reactions and nonlinear couplings among multiple rates may likewise modify the pre-explosive composition. Extending the present temperature-resolved approach to a broader range of progenitor models and reaction networks will help establish how widely the sensitivity around $T\simeq0.25\,\mathrm{GK}$ applies across the PISN progenitor population. Within the models considered here, our results identify this temperature range as particularly sensitive to reaction-rate variations, highlighting the value of temperature-resolved analyses for understanding how nuclear reaction-rate uncertainties affect PISN outcomes.

\section*{Funding}
This work is supported by JSPS KAKENHI grant Nos. JP24H00008, JP24H02245, JP24K00668, JP25H01273, JP25K01035, JP26K00723, and JP26KJ0832. HK was supported by the RIKEN Junior Research Associate Program.

\begin{ack}
HK thanks Masaaki Kimura for fruitful discussions.
\end{ack}

\section*{Data availability}
The data underlying this article will be shared upon reasonable request to the corresponding author.

\bibliographystyle{pasjedit}
\bibliography{bib}

\end{document}